# Superconductivity at 52 K in hydrogen-substituted LaFeAsO$_{1-x}$H$_x$ under high pressure

H. Takahashi [1,*], H. Soeda[1], M. Nukii[1], C. Kawashima[1], T. Nakanishi[2], S. Iimura[3], Y. Muraba[4], S. Matsuishi[4], and H. Hosono[3,4]

[1] Department of Physics, College of Humanities and Sciences, Nihon University, Tokyo, Japan
[2] Department of Physics, School of Science and Technology for Future Life, Tokyo Denki University, Tokyo, Japan
[3] Materials and Structures Laboratory, Tokyo Institute of Technology, Yokohama, Japan
[4] Materials Research Center for Element Strategy, Tokyo Institute of Technology, Yokohama, Japan

* *hiroki@chs.nihon-u.ac.jp*

The 1111-type iron-based superconductor $Ln$FeAsO$_{1-x}$F$_x$ ($Ln$ stands for lanthanide) is the first material with a $T_c$ above 50 K, other than cuprate superconductors. Electron doping into LaFeAsO by H, rather than F, revealed a double-dome-shaped $T_c$-$x$ diagram, with a first dome (SC1, 0.05<x<0.20) and a second dome (SC2, 0.2<x<0.5). Here, we report the $T_c$ for the whole hydrogen-doping range in LaFeAsO$_{1-x}$H$_x$ under pressures of up to 19 GPa. $T_c$ rises to 52 K at 6 GPa for the $T_c$-valley composition between the two $T_c$ domes. This is the first instance of the $T_c$ exceeding 50 K in La-1111-type iron-based superconductors. On the other hand, the $T_c$ of SmFeAsO$_{1-x}$H$_x$ decreased continually, keeping its single-dome structure up to 15 GPa. The present findings strongly suggest that the main reason for realization of the $T_c$ >50 K observed in RE-1111 compounds (RE: Pr, Sm, and Gd) at ambient pressure is the merging of SC1 and SC2.

The discovery of the iron-based superconductor LaFeAsO$_{1-x}$F$_x$ had a significant impact on condensed matter physics as a new platform for studying high-$T_c$ superconductivity[1]. Several kinds of materials have been developed, and the highest $T_c$ obtained to date is 58.1 K for $Ln$-1111-type $Ln$FeAsO$_{1-x}$F$_x$ ($Ln$ stands for lanthanide)[2]. The undoped compound LaFeAsO is a Pauli paramagnetic metal with a tetragonal symmetry at room temperature, and undergoes structural (tetragonal-to-orthorhombic) and magnetic (paramagnetic-to-antiferromagnetic) transitions[3,4]. Superconductivity appears when both transitions are suppressed with carrier doping, The interplay between superconductivity and magnetism is important, and several models have been proposed for the appearance of superconducting phase next to the antiferromagnetic phase. For lightly-doped materials, a spin fluctuation resulting from the Fermi surface nesting between hole and electron pockets is a plausible candidate for the pairing mechanism[5,6].

Hydrogen-doped $Ln$-1111 materials were successfully synthesized using the high solubility limit of hydrogen[7], covering the over-doped region, which has never been studied before because of the low solubility limit of fluorine. A combined study of neutron diffraction measurements with density functional calculations on $Ln$FeAsO$_{1-x}$D$_x$ demonstrated that these hydrogen atoms are incorporated as H$^-$ ions at the O$^{2-}$ sites. The hydrogen doping causes superconductivity as the same manner as fluorine doping for Sm[8], Ce[9] and La-1111[7], as shown in the previous studies. A complete single $T_c$ dome is observed in Ce and Sm-1111[7], while a double-dome-shaped $T_c$ curve is obtained in LaFeAsO$_{1-x}$H$_x$, which has hitherto not been observed for $Ln$-1111. The phase diagrams of LaFeAsO$_{1-x}$H$_x$ and SmFeAsO$_{1-x}$H$_x$ at ambient pressure are presented in Fig. 1. The $T_c$ curve of the first dome (SC1) in the lightly-doped region almost coincides with the fluorine-doped La-1111. However, the second dome (SC2) in the heavily-doped region has the maximum $T_c$ of 36 K at $x = 0.36$, which is higher than that of SC1. The SC2 is far from the magnetic phase of x=0. This finding suggests that the rigid-band picture does not hold in the highly-doped region[7,10] because the size and shape of hole pocket on the Fermi surface differs fairly from those of the electron pockets. Thus, orbital fluctuation and/or spin fluctuation resulting from another have been proposed recently as the superconductivity mechanism in the highly-doped region[10]. In addition, a new antiferromagnetic phase was recently discovered by NMR[11] and neutron diffraction measurements[12] in the over-doped region (x=~0.5) in LaFeAsO$_{1-x}$H$_x$, where the $T_c$ of the SC2 vanishes. This magnetic phase is expected to be the key to understanding the appearance of the second $T_c$ dome.

Applying pressure is a very effective means of examining the properties of such layered superconductors. The use of pressure allows the electronic structure to be modified by a physical means, without inducing disorders and/or impurities. A number of important results have been obtained for iron-based superconductors using high-pressure techniques. Applying a pressure of 4 GPa on the first discovered iron-based superconductor LaFeAsO$_{1-x}$F$_x$ raised the $T_c$ substantially, from 26 K to 43 K[13]. This large pressure effect is recognized as one of the striking features of iron-based superconductors. In addition to the enhancement of $T_c$, pressure-induced superconductivity has been discovered in, for example, LaFeAsO[14] and SrFe$_2$As$_2$[15], and a pressure-induced higher-$T_c$ phase has been identified in 11-type iron chalcogenide FeSe[16]. These high-pressure results can provide guiding principles for the development of new superconductors and pave the way to promising new means of investigating iron-based superconductors. High-pressure experiments have already been performed on LaFeAsO$_{1-x}$H$_x$ at up to 3 GPa, revealing that the double $T_c$ dome (domes SC1 and SC2) merged into a single dome at 3 GPa[7]; however, it remains unclear how this merge occurred. It should also be intriguing to examine what effects are induced by further compression (> 3 GPa).

Here, we study high-pressure effects over the entire hydrogen-doping range in LaFeAsO$_{1-x}$H$_x$ under pressures of up to 19 GPa. The $T_c$ at the $T_c$-valley composition between the two $T_c$ domes is greatly increased, from 18 K to 52 K, under a pressure of 6 GPa. This is the first known instance of 50 K being exceeded in a La-1111-type iron-based superconductor. Such a large enhancement in $T_c$ suggests that the two factors giving rise to SC1 and SC2 are effectively merged at the $T_c$-valley composition, leading to a higher $T_c$ comparable to those of high-$T_c$ $Ln$-1111 compounds ($Ln$ = Nd, Sm, and Gd). These results are compared with the superconducting properties of SmFeAsO$_{1-x}$H$_x$ under high pressure.

**Results**
**LaFeAsO$_{1-x}$H$_x$.** Figure 2a shows the temperature dependence of electrical resistance for LaFeAsO$_{1-x}$H$_x$ with $x$ = 0.18 ($T_c$-valley composition) at each pressure obtained using a diamond anvil cell (DAC). The sudden decrease in resistance in the low temperature range is regarded to be due to the superconducting transition. The onset $T_c$ is given by the intersection of two extrapolated lines, one drawn through the resistance curve for the normal state just above $T_c$, and the other drawn through the steepest part of the resistance curve for the superconducting state, as shown in Fig. 2a. The pressure dependence of the $T_c$'s for LaFeAsO$_{1-x}$H$_x$ with $x$ = 0.07, 0.18, 0.30, and 0.44 is plotted in Fig. 2b. The $T_c$'s obtained using a piston-cylinder cell below 2.5 GPa are shown in the same figure. The pressure dependence of $T_c$ for $x$ = 0.18 exhibits a dome shape with the maximum $T_c$ of 52 K at 6 GPa. The $T_c$ vs. $P$ curves for the other compositions, with the exception of $x$ = 0.44, also show a dome-shaped pressure dependence. The $T_c$ of the heavily-doped LaFeAsO$_{1-x}$H$_x$ with $x$ = 0.44 decreases monotonically with applied pressure. Figure 2d shows the pressure dependence of normalized lattice constants $a$ and $c$ at 60 K and the atomic position of As ($z_{As}$) estimated by density-functional-theory (DFT) calculations. Although a small discrepancy between the calculated and experimental $z_{As}$ has been reported previously due to the strong electronic correlations[17], it is enough to see the trend of the pressure dependency. It is clear that as the pressure increased, the lattice is compressed and the As-Fe-As angle (α) given by $2\arctan\{a/c \times 1/[2(z_{As} - 0.5)]\}$, decreases monotonically.

**SmFeAsO$_{1-x}$H$_x$.** Figure 3a shows the temperature dependence of electrical resistivity for lightly-doped SmFeAsO$_{1-x}$H$_x$ with $x$ = 0.03 at each pressure obtained using a piston-cylinder cell. The large change in resistivity at around 100 K corresponds to structural and magnetic transitions similar to those reported for LaFeAsO. This transition temperature $T_0$ is given by the temperature at which d$\rho$/d$T$ shows its peak value and is suppressed at a rate of -10.0 K/GPa under pressure. On the other hand, a $T_c$ is also observed at around 5 K that increases at a rate of +1.2 K/GPa. The pressure dependence of $T_c$ for SmFeAsO$_{1-x}$H$_x$ with $x$ = 0.03, 0.07, 0.10, 0.20, 0.32, 0.34, and 0.38 is shown in Fig 3b. Except for $x$ = 0.03, $T_c$ decreases with increasing applied pressure, at rates ranging from -1.2 to -0.6 K/GPa. High-pressure resistance measurements using a DAC are carried out for $x$ = 0.20 and 0.38 at up to 15 GPa, in which $T_c$ decreases at the same rate as in the pressure region below 2.5 GPa.

**Discussion**
In layered materials, the anisotropic contraction usually induces a significant change in the charge distribution, leading to a change in the electronic state. For iron-based superconductors, such a effect may be expected because of the layered structure. Moreover, experimental work has indicated that electronic states are sensitive to the local structure around iron, such as the bond angle and bond length. For 1111-type superconductors, it is widely believed that $T_c$ increases as the FeAs$_4$ tetrahedron approaches its regular shape having an As-Fe-As bond angle $\alpha$ of 109.47° (so-called

"Lee's plot")[18]. According to this idea, spin fluctuation is the key ingredient in the glue of the electron pair, since the calculated band structure for the crystal structure having a regular tetrahedron is suitable for the development of spin fluctuation resulting from the Fermi surface nesting between hole and electron pockets.

Figure 4a shows the $T_c$–$x$ phase diagrams for several pressures up to 13 GPa. Under an applied pressure of 1 GPa, the $T_c$ domes SC1 and SC2 are enhanced, and SC2 shifts to the lightly-doped side, while SC1 shifts in the opposite direction. Increasing the applied pressure to 6 GPa greatly raises the $T_c$-valley, which causes the double $T_c$ dome to merge into a single $T_c$ dome and to shift to the lightly-doped side as a single $T_c$ dome. The width of this single $T_c$ dome is smaller than the double $T_c$ dome. For hydrogen-doped La-1111 and Sm-1111 (see Fig. 1a), the optimal $x$ indicated by the arrows shifts to the lightly-doped side in the order La to Sm, and the width of the $T_c$ dome is larger for La-1111 than for Sm-1111. However, these features are widely recognized in hydrogen-doped $Ln$-1111($Ln$ = La, Ce, Sm, and Gd), as the ionic radius of $Ln$ decreases in the order La > Ce > Sm > Gd[7]. Since the crystal lattice is compressed under pressure, it is reasonable to consider that the double dome $T_c$ in La-1111 would deform and approach the shape of the Sm-1111 single $T_c$ dome under high pressure. Thus, it is plausible that the dramatic rise in $T_c$ to 52 K, observed in the $T_c$-valley, is attributable to the combined effect of pressure on the two $T_c$ domes. For hydrogen-doped Sm-, Ce-, and Gd-1111, the observed single $T_c$ dome having an optimum $T_c$ above 40 K is presumably a consequence of the effective merge of the two $T_c$ domes. In the case of La-1111, increasing the applied pressure to 13 GPa leads to further shrinkage of the $T_c$ dome. Figure 4b shows the $T_c$–$x$ phase diagram of SmFeAsO$_{1-x}$H$_x$ for a number of pressures up to 15 GPa. $T_c$ decreases monotonically with pressure, and the single $T_c$ dome shrinks and shifts to the lightly-doped side, as in the case of La-1111 above 6 GPa.

For LaFeAsO$_{1-x}$H$_x$, the power law exponent $n$ of the normal-state electrical resistivity is ~2 in the SC1 region, which reflects the Fermi liquid-like properties. On the other hand, in the SC2 region, which has a higher $T_c$, a non-Fermi liquid-like behavior ($n \approx 1$) was observed. It was indicated that $T_c$ increases as $n$ approaches unity[7]. A $T_c$ above 40 K has also been observed in other hydrogen-doped $Ln$-1111 for $n \approx 1$. However, for LaFeAsO$_{1-x}$H$_x$ at the $T_c$-valley composition, $n$ does not change significantly with increasing pressure, despite the dramatic change in $T_c$ from 16 to 52 K under an applied pressure of 6 GPa. Figure 2c shows the pressure dependence of $n$ for the $T_c$-valley material. The $n$ value decreases very gradually from 2.0 at ambient pressure to unity. The correlation between $T_c$ and $n$ observed at ambient pressure is largely absent in this high-pressure case. This result suggests that under high pressure, the change in the electronic state of LaFeAsO$_{1-x}$H$_x$ is insignificant, while the change in its $T_c$ is large.

Structurally, it is reasonable to consider that under high pressure, the FeAs$_4$ tetrahedron deforms, stretching in the inter-layer direction. Since the inter-layer bond is not as strong as the covalent Fe-As bond, the former is more susceptible to pressure. Thus, the FeAs$_4$ tetrahedron stretches in the inter-layer direction when it deforms, as demonstrated in Ba(Fe$_{1-x}$Co$_x$)$_2$As$_2$[20]. This deformation decreases the As-Fe-As bond angle $\alpha$. In the case of LaFeAsO$_{1-x}$H$_x$, the decrease of bond angle $\alpha$ with increasing pressure is demonstrated by x-ray diffraction measurements and DFT calculations, as shown in Fig. 2d. These results indicate that hydrogen doping and pressure both cause the tetrahedron to deform and approach a regular tetrahedron. It is revealed that the structure of La1111 approaches to Sm1111 with applying pressure, which is consistent with the pressure effect on $T_c$.

The interplay between magnetism and superconductivity in high-$T_c$ cuprates and heavy-fermion materials has been examined since before the discovery of iron-based superconductors. For LaFeAsO$_{1-x}$H$_x$, NMR experiments[11] and theoretical calculations[10] indicate that spin and orbital fluctuations develop around the second magnetic ordering phase in the over-doped region. Because theoretical calculations suggest that the spin and orbital fluctuations develop mutually in this system in the lightly- and heavily-doped regions[10], we believe that the large enhancement in $T_c$ to 52 K at the $T_c$-valley composition is caused by combined effects arising from both SC1 and SC2. Our preliminary measurements show that the second magnetic ordering is suppressed under an applied pressure. We also note that the $T_c$ in the heavily-doped region is suppressed by pressure, while the $T_c$ in the lightly-doped region is enhanced by pressure, when the undoped magnetic phase is suppressed. This suggests that a different relationship exists between the magnetic phase and superconductivity for the two $T_c$ domes.

Recent theoretical calculations have reproduced the double-dome $T_c$ behavior in LaFeAsO$_{1-x}$H$_x$ by considering two kinds of pairing causes for SC1 and SC2[21,22]. In terms of orbital fluctuation[21], the approximate $s_{++}$-wave gap structures due to orbital fluctuations were obtained for both the undoped and over-doped extremes and the double-dome $T_c$ was obtained by switching the dominant quadrupole fluctuation from the SC1 phase (conventional nematic orbital fluctuation) to the SC2 phase (non-nematic orbital fluctuation). On the other hand, the spin fluctuation gives rise to the $s_{+-}$-wave gap structure. By extending these theoretical considerations, the double-dome $T_c$ curve could be explained through the relationship between next-nearest-neighbor hoppings ($t_1$) between the $d_{xy}$ orbital of iron sites and second-nearest-neighbor hoppings ($t_2$)[22]. For SC1, $t_1 > t_2$ is established, and the spin fluctuation arising from the Fermi surface nesting between electron and hole pockets is thought to play an important role in the superconducting mechanism. By contrast, for SC2, $t_1 < t_2$ and spin fluctuation due to a different cause is thought to be important, leading to the higher $T_c$ maximum. For other hydrogen-doped $Ln$-1111 materials, a double-dome $T_c$ has not been observed for $T_c$ higher than 40K (SmFeAs$_{1-x}$P$_x$O$_{1-y}$H$_y$ system has a double dome structure but their optimal $T_c$=~20 K[26])., because $t_1 < t_2$ occurs in the more lightly-doped region. Sm-1111 has a $T_c$ of 56 K at the optimal hydrogen concentration, where the $t_1 < t_2$ condition is thought to be stable. Therefore, we may conclude that the effect of pressure on $T_c$ in Sm-1111 is not as large as it is in La-1111; in La-1111, the two hopping components compete with each other around the $T_c$-valley composition and are thus sensitive to pressure.

In summary, we measure the pressure dependence of $T_c$ for the whole hydrogen-doping range in LaFeAsO1$_{-x}$H$_x$ under high pressure. The $T_c$ dome of SC2 is enhanced and shifts to the lightly-doped side upon applying pressure. Judging from the phase diagram obtained under high pressure, the SC2 phase becomes dominant in the high-pressure phase. We believe that the enhancement in $T_c$ is caused by a combination of multiple effects. In particular, the two factors giving rise to SC1 and SC2 are effectively merged at the $T_c$-valley composition, yielding a higher $T_c$, comparable to those of high-$T_c$ $Ln$-1111 compounds. The $T_c(x)$ curve is suppressed above 6 GPa, as seen in the phase diagram, similarly to what is observed in the Sm-1111 phase diagram under high pressure. For SmFeAsO$_{1-x}$H$_x$, the two phases of SC1 and SC2 are thought to merge at ambient pressure. High-pressure studies of the magnetic phase on the highly-doped side are important to examine the interplay between superconductivity and magnetism.

**Methods**

 **Resistivity measurements under high pressure**

Electrical resistivity measurements under high pressure were performed by the standard dc four-probe method. Pressures of up to 2.5 GPa were applied at room temperature using a WC piston and NiCrAl cylinder device. A liquid pressure-transmitting medium (Daphne oil 7474) was used to maintain hydrostatic conditions. A diamond anvil cell (DAC) made of CuBe alloy was used for electrical resistance measurements at pressures up to 30 GPa: in this case, the sample chamber comprising a rhenium gasket was filled with powdered NaCl as the pressure-transmitting medium, and thin (10-µm-thick) platinum ribbons were inserted into the sample chamber to act as leads for the standard dc four-probe analysis. The dimensions of the samples were 0.1×0.1×0.03 mm. A thin BN layer acted as electric insulation between the leads and the rhenium gasket. Fine ruby powder scattered throughout the sample chamber was used to determine the pressure by the standard ruby fluorescence method. The lack of a measurement of zero resistance in La1111 as shown in Fig. 2a could be due to technical limitations inherent in the experimental apparatus, since fully symmetric hydrostatic compressive stress could not be applied inside the DAC when using a solid pressure-transmitting medium. On the other hand, the resistivity measurement of Sm1111 shows zero resistivity as shown in Fig. 3a, since it was carried out using piston cylinder apparatus, in which liquid pressure-transmitting medium was used to maintain hydrostatic conditions.

 **Density functional theory calculations**

Non-spin-polarized DFT calculations for $LaFeAsO_{0.82}H_{0.18}$ were performed using the projected augmented plane-wave method[23] implemented in the Vienna ab initio simulation program (VASP) code [24] and the generalized gradient approximation Perdew–Burke–Ernzerhof functional for solid[25]. Experimental lattice-constants determined under high pressure were used and plane-wave basis-set cutoff was set to 600 eV. The 10 × 10 × 6 meshes were taken for the Brillouin zone integration. For doping-effect, the virtual crystal approximation was used by replacing the oxygen potential with the 0.82:0.18 mixture of oxygen and fluorine potentials. Atomic positions under high pressure were calculated by the structure relaxation with fixed lattice constants.

**Acknowledgements**
This work was partly supported by JSPS Grants-in-Aid for Scientific Research (B) (24340088) and the Strategic Research Base Development Program for Private Universities (2009, S0901022) of MEXT. The research at Tokyo Tech was supported by the MEXT Element Strategy Initiative to form a Core Centre and JSPS FIRST Project.

**Author information**
Reprints and permissions information are available at www.nature.com/reprints. Correspondence and requests for materials should be addressed to H.T.(hiroki@chs.nihon-u.ac.jp).

**Author contributions**
H.T. and H.H. planned the research. S.I., Y.M. and S.M. performed the high pressure synthesis. H.S.,N.N., C.K. and T.N. carried out high-pressure measurement.   S.I. and S.M peformed DFT calculations. H.T., H.H., S.I. and H.S. discussed the results and wrote the manuscript.

**Competing financial interests**
The authors declare no competing financial interests.

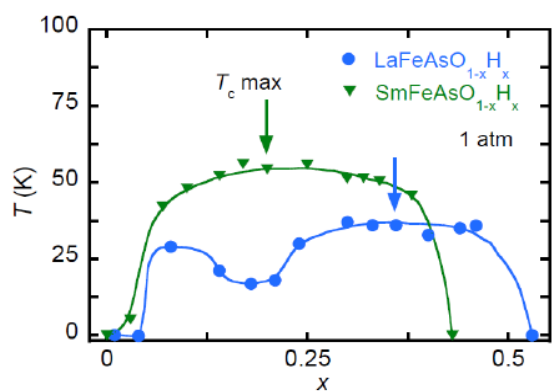

Figure l. **Phase diagram in hydrogen-doped 1111 materials**
Superconductive phase diagrams for $Ln$FeAsO$_{1-x}$H$_x$ ($Ln$ =La and Sm). A double-dome $T_c(x)$ is observed in LaFeAsO$_{1-x}$H$_x$[7]. The two kinds of superconducting phases, SC1 and SC2, are thought to have different origins. The arrows show the maximum $T_c$, which shifts to the lightly-doped side in the order La to Sm. This shift in $T_c$-dome is observed for $Ln$FeAsO$_{1-x}$H$_x$ ($Ln$ =La, Ce, Sm, and Gd) in the order La to Gd, i.e., in order of decreasing ionic radius.

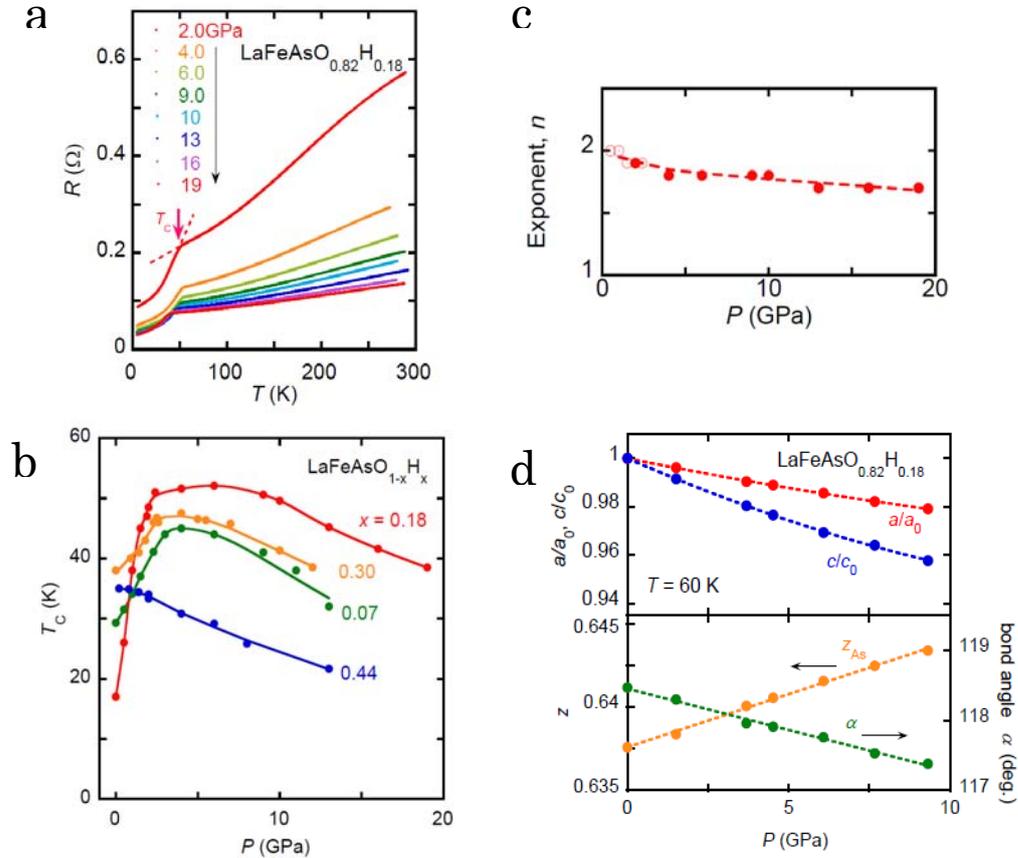

Figure 2. **Superconducting and structural properties in LaFeAsO$_{1-x}$H$_x$ under high pressure**

**a)** Temperature dependence of the electrical resistance for $x = 0.18$ using DAC. The onset $T_c$ is determined to be the intersection of two extrapolated lines, one drawn just above $T_c$ on the resistance curve in the normal state and the other drawn through the steepest part of the resistance curve in the superconducting state. These extrapolated lines are shown on the data for 2 GPa. The superconducting transition is clearly observed for each measurement, although the zero resistance is not observed down to 4 K. **b)** Pressure dependence of $T_c$ for $x = 0.07$, 0.18, 0.30, and 0.44. The $T_c$ is defined as the onset temperature of the transition. Solid curves are a guide for the eye. **c)** Pressure dependence of $n$ for $x = 0.18$. The $n$ value decreases slightly under high pressure. **d)** Pressure dependence of lattice constants normalized to the ambient pressure values, $z_{As}$ and As-Fe-As angle $\alpha$. The lattice constants are obtained by high-pressure X-ray diffraction, and the atomic positions are estimated by DFT calculations using the experimental lattice constants.

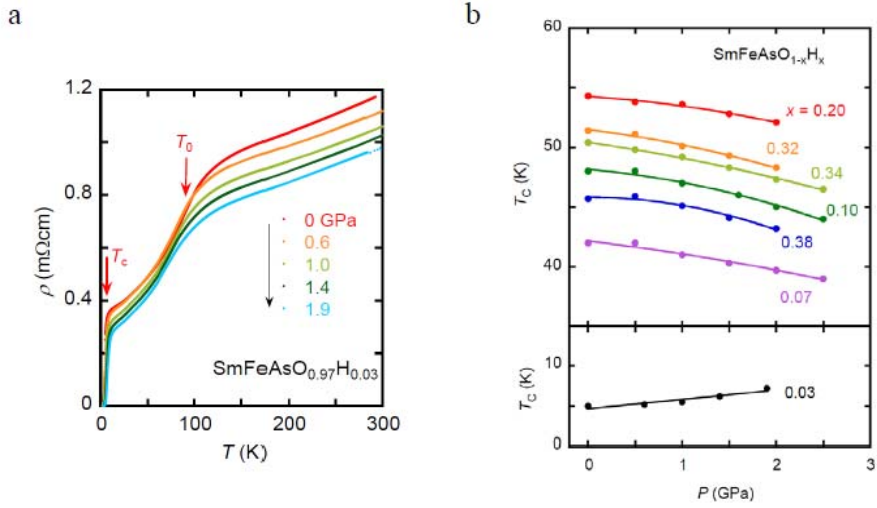

Figure 3. **Superconducting properties in SmFeAsO$_{1-x}$H$_x$ under high pressure**
a) Temperature dependence of electrical resistivity for $x$ = 0.03, obtained using the piston-cylinder device. The anomalous decrease in resistivity observed around 100 K corresponds to structural and magnetic transitions. The transition temperature $T_0$ is determined as the temperature that shows the peak d$\rho$/d$T$ value and decreases with increasing pressure with an initial slope of -10.0 K/GPa. It is indicated by arrows on the data for 0 GPa. The superconducting transition temperature is also observed at around 5 K and increases with pressure. b) Pressure dependence of $T_c$ for $x$ = 0.03, 0.07, 0.10, 0.20, 0.32, 0.34, and 0.38. The $T_c$ is defined as the onset temperature of the transition. Solid curves are a guide for the eye.

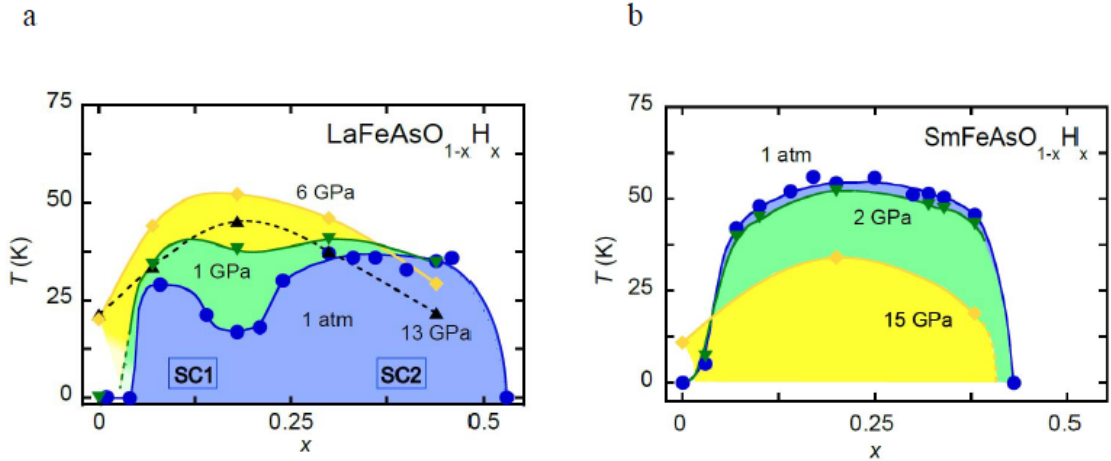

Figure 4. **Phase diagram in LaFeAsO$_{1-x}$H$_x$ and SmFeAsO$_{1-x}$H$_x$ under high pressure**
a) The phase diagrams of LaFeAsO$_{1-x}$H$_x$ under the pressures 1 atm and 1, 6, and 13 GPa. As the pressure is increased to 6 GPa, the $T_c$-valley is greatly raised and the $T_c$ dome SC2 shifts to the lightly-doped side. A large enhancement in $T_c$ to 52 K is observed at the $T_c$-valley. The $T_c$ dome is suppressed above 6 GPa. b) The phase diagrams of SmFeAsO$_{1-x}$H$_x$ for the pressure of 1 atm and 2 and 15 GPa. $T_c$ is suppressed upon the application of pressure, except in the case of the undoped material. Pressure-induced superconductivity has previously been identified in undoped SmFeAsO by our group[19]. The single $T_c$ dome seems to shift to the lightly-doped side under high pressure. The same trend is seen in the case of LaFeAsO$_{1-x}$H$_x$.